\documentclass[12pt]{article}
\usepackage{amsmath}
\usepackage{mitpress}

\newdimen\dummy
\dummy=\oddsidemargin
\input{tcilatex}

\begin{document}

\title{Some consequences of the Einstein covariance principle in electrodynamics of
media}
\author{\textbf{Vahram M. Mekhitarian and Vanik E. Mkrtchian} \\
Institute for Physical Research, Armenian Academy of Sciences, Ashtarak-2,
378410, Republic of Armenia.}
\maketitle

\begin{abstract}
We show that the Einstein covariance principle provides an opportunity to
generate infinitely many solutions of given covariant equation from a known
one. With use of this statement we derive exact expressions for charge and
current densities in a medium in an arbitrary external field. We obtain that
in a homogeneous nondisipersive medium the field of a point particle differs
at short-distances from the well known expression. We also demonstrate that
in a linear medium second harmonic generation and optical detection become
possible in the field of radially polarized radiation..

\textbf{Key words \ }General relativity\textbf{, }covariance principle,
charge and current densities.

\end{abstract}

\section{\protect\bigskip Introduction}

The requirement of general covariance of the equations describing different
processes in the nature is one of the corner-stones of Einstein general
relativity \cite{ein} and has enormous significance in modern theoretical
physics. When combined with the equivalence principle it reduces the
gravitation to the metric properties of space-time and shows interconnection
between the geometry of space-time continuum and the material processes.

In this article we make an attempt to discover another powerful aspect of
the Einstein covariance principle. Namely, we are going to show that, the
covariance principle provides new methods (algorithms) for solution of
different problems of theoretical and mathematical physics.

In Sec.2 we show that with use of the covariance principle it is possible to
generate infinitely many solutions of given covariant equation from a known
one. The new, generated, solutions correspond to physical situations
differing (in boundary conditions, external fields) from those in the known
solution. For instance, having a free solution of the given covariant
equation, we can generate solutions of the same equation in the external
field.

We succeeded in Sec.3 to perform this procedure for the continuity equation
for an ensemble of charged point particles and in Sec.4 for a continuous
medium using the Euler transformation i.e. with the help of Euler
transformation we can pick up the solutions of the continuity equation in
the external field from the infinite set of the new solutions of continuity
equation.

As a result, in Sec.4 we obtain exact expressions for the charge and current
densities in a medium in the presence of an arbitrary external field. These
formulae contain the external field in implicit way via Euler displacement
field, which provides an opportunity to define polarization of the medium in
natural way independent of the kind of the medium. Then, we show that in the
case of homogeneous isotropic media it is possible to unambiguously
determine the polarization and magnetization vectors in terms of the medium
polarization.

In the Secs.5 and 6 we discuss some physical consequences of the results of
Sec.4 considering two "textbook" problems of the linear electrodynamics.

In Sec.5 we solve the problem of the field of a charged point particle in a
homogeneous dielectric medium and find out that in this case the electric
field is determined by a cubic algebraic equation. Analyzing solutions of
this equation we detain a deviation from, the well known result for small
distances from the point charge.

In Sec.6 we consider propagation of radially polarized radiation in a
homogeneous linear dielectric medium. Our calculations show that in this
case (linear medium) it is possible to observe nonlinear phenomena like
optical detection and second harmonic generation.

Sec.7 concludes the paper with some remarks.

\section{The main statement}

The Einstein covariance principle \cite{ein} claims that any physical law
must have a covariant form i.e. if physical quantities $A,B,...$according to
some physical law, are related by an equation

\begin{equation}
F\left( A\left( X\right) ,\text{ }\hat{L}\left[ B\left( X\right) \right]
,...\right) =0,  \tag{1.a}
\end{equation}
in coordinates $X$ ( $\hat{L}$ is an operator, differential, integral,
etc.), the functional relation should be the same in any other coordinates $%
X^{\prime }$i.e. 
\begin{equation}
F\left( A^{\prime }\left( X^{\prime }\right) ,\text{ }\hat{L}^{\prime }\left[
B^{\prime }\left( X^{\prime }\right) \right] ,...\right) =0.  \tag{1.b}
\end{equation}

But, each physical quantity has certain transformation property (scalar,
vector, tensor etc.), i.e., for\ a given coordinate transformation 
\begin{equation}
X^{\prime i}=W^{i}\left( X\right) ,\text{ }(i=0,1,2,3)  \tag{2}
\end{equation}
these quantities are transformed as 
\begin{equation}
A\left( X\right) =\hat{\Lambda}\left( X\right) A^{\prime }\left( W\left(
X\right) \right) ,\text{ }B\left( X\right) =\hat{\Lambda}\left( X\right)
B^{\prime }\left( W\left( X\right) \right) ,...  \tag{3}
\end{equation}
where $\hat{\Lambda}=1$ for scalar quantities, for the vectors $\hat{\Lambda}
$ is a matrix (see.(8)) , for higher rank tensors $\hat{\Lambda}$ is a
direct product of matrices. Insertion of (3) into (1.a) gives the equation 
\begin{equation}
F\left( \hat{\Lambda}\left( X\right) A^{\prime }\left( W\left( X\right)
\right) ,\text{ }\hat{L}\left[ \hat{\Lambda}\left( X\right) B^{\prime
}\left( W\left( X\right) \right) \right] ,...\right) =0,  \tag{4}
\end{equation}
which, by comparing with (1.b), leads to the following statement:

\textit{If }$A\left( X\right) ,$\textit{\ }$B\left( X\right) ,...$\textit{\
satisfy the covariant equation}

\begin{equation}
F\left( A\left( X\right) ,\text{ }\hat{L}\left[ B\left( X\right) \right]
,...\right) =0,  \tag{5.a}
\end{equation}
\textit{\ then} 
\begin{equation}
\hat{\Lambda}\left( X\right) A\left( W\left( X\right) \right) ,\text{ \ }%
\hat{\Lambda}\left( X\right) B\left( W\left( X\right) \right) ,...  \tag{5.b}
\end{equation}
\textit{satisfy the same equation for any }$W^{i}\left( X\right) $\textit{\ }%
$\left( i=0,1,2,3\right) .$

Thus, Einstein covariance principle provides an opportunity to generate
infinite many new solutions for a\ given covariant equation from a known one
by means of transformations within the frames of general relativity. The
kind of covariant equation is not important for this statement: the equation
may be linear or non-linear, differential or integro-differential, etc..

\section{\protect\bigskip Continuity equation}

As an application of the statement above let us consider the covariant
continuity equation in electrodynamics

\begin{equation}
\frac{1}{\sqrt{-g\left( X\right) }}\partial _{i}\left( \sqrt{-g\left(
X\right) }j^{i}\left( X\right) \right) =0  \tag{6}
\end{equation}
where $-g\left( X\right) $ is the determinant of metric tensor ( we use
notations of the well known Landau-Lifshitz book \cite{lan1}). It follows
from (5) that, for any transformation $W^{i}\left( X\right) ,$ 
\begin{equation}
\hat{\Lambda}\sqrt{-g}\equiv \left\Vert \Lambda \left( X\right) \right\Vert 
\sqrt{-g\left( W\left( X\right) \right) },\ \hat{\Lambda}j^{i}\equiv \tilde{%
\Lambda}_{n}^{i}\left( X\right) j^{n}\left( W\left( X\right) \right) , 
\tag{7}
\end{equation}
satisfies the Eq.(6). Here we took into account the four-vector character of
the current and the transformation law for the determinant of metric tensor: 
\begin{equation}
\Lambda _{j}^{i}(X)\equiv \partial _{j}W^{i}(X),\text{ }\tilde{\Lambda}%
_{n}^{i}\Lambda _{j}^{n}=\delta _{j}^{i}.  \tag{8}
\end{equation}

Writing (6), (7) in the Cartesian coordinates, we now claim that if $%
j_{0}^{i}\left( X\right) $ is a solution of 
\begin{equation}
\partial _{i}j_{0}^{i}=0,  \tag{9}
\end{equation}
then 
\begin{equation}
j^{i}\left( X\right) \equiv \left\Vert \Lambda \left( X\right) \right\Vert 
\tilde{\Lambda}_{n}^{i}\left( X\right) j_{0}^{n}\left( W\left( X\right)
\right)  \tag{10}
\end{equation}
satisfies the same equation $\left( 9\right) $ for any transformations $%
W^{i}\left( X\right) $.

The expression $\left( 10\right) $ has very important consequences in the
electrodynamics. It provides a possibility to express charge and current
densities in the arbitrary external field via charge and current densities
of the undisturbed system.

This suggestion may be proved for an ensemble of charged point particles
without using the statement (5). Indeed, let us consider an ensemble of
particles having charge $e_{a}$ and trajectories $\mathbf{r}_{a}^{0}\left(
t\right) $ $\left( a=1,2,...\right) $ in the absence of the external field.
In this case the current four-vector at the point $X^{\prime }\equiv \left(
ct^{\prime },\mathbf{r}^{\prime }\right) $ is given by \cite{lan1} 
\begin{equation}
j_{0}^{i}\left( X^{\prime }\right) =c\dsum\limits_{a}e_{a}\delta \left( 
\mathbf{r}^{\prime }-\mathbf{r}_{a}^{0}\left( t^{\prime }\right) \right) 
\frac{dX^{\prime i}}{dX^{\prime 0}}.  \tag{11.a}
\end{equation}
\ 

In the presence of an external field the trajectories of particles are
changed $\mathbf{r}_{a}^{0}\left( t\right) \rightarrow \mathbf{r}_{a}\left(
t\right) $ and the current at the point $X\equiv \left( ct,\mathbf{r}\right) 
$\ becomes 
\begin{equation}
j^{i}\left( X\right) =c\dsum\limits_{a}e_{a}\delta \left( \mathbf{r}-\mathbf{%
r}_{a}\left( t\right) \right) \frac{dX^{i}}{dX^{0}}.  \tag{11.b}
\end{equation}
\ 

\bigskip Now, if we perform the Euler transformation (see Appendix) 
\begin{equation}
X^{\prime i}=X^{i}-U^{i}\left( X\right) ,\text{ \ }U^{i}=\left( 0,\mathbf{u}%
\left( \mathbf{r,}t\right) \right)  \tag{12}
\end{equation}
in $\left( 11.a\right) $ and use the well-known formula 
\begin{equation*}
\dprod\limits_{i=1}^{n}\delta \left( x_{i}-\alpha _{i}\right) =\frac{1}{%
\left\vert J\right\vert }\dprod\limits_{i=1}^{n}\delta \left( \xi _{i}-\beta
_{i}\right) \text{,\ \ \ }J\equiv \frac{\partial \left(
x_{1}....x_{n}\right) }{\partial \left( \xi _{1}.....\xi _{n}\right) },
\end{equation*}
we arrive at the expression 
\begin{equation}
j^{i}\left( X\right) \equiv \left\Vert \Lambda \left( X\right) \right\Vert 
\tilde{\Lambda}_{n}^{i}\left( X\right) j_{0}^{n}\left( X-U\right) ,  \tag{13}
\end{equation}
which coincides with $\left( 11.b\right) $ under the condition 
\begin{equation}
\mathbf{u}\left( \mathbf{r}_{a}\left( t\right) \mathbf{,}t\right) =\mathbf{r}%
_{a}\left( t\right) \mathbf{-r}_{a}^{0}\left( t\right) .  \tag{14}
\end{equation}
\ \ 

\ This means that in an arbitrary external field the current four-vector $%
j^{i}\left( X\right) $ is expressed linearly in terms of the undisturbed
current four-current $j_{0}^{i}$ at the point $X-U$.

As we see, $\mathbf{u}(\mathbf{r,}t)$ is a field which equals, on the
trajectories of particles, to the displacement of these trajectories caused
by external forces and, hence, it has the similar meaning as in the theory
of elasticity \cite{lan2}.

On the other hand, the expression $\left( 13\right) $ is a special case of $%
\left( 10\right) $ when $W^{i}\left( X\right) $ is the Euler transformation $%
\left( 12\right) $ with the additional condition $\left( 14\right) $. Hence,
we can say that, by using the covariance principle, we are able to connect
current and charge densities of disturbed and undisturbed system of an
ensemble of point particles in any external field with the help of the Euler
transformation. This suggestion is rather general. In Sec.4 we will come to
the same conclusion for the four-current in a medium without referring to
point structure of the particles in medium.

\section{ Charge and current densities in a medium}

The expression (10) for the current provides an opportunity to reformulate
unambiguously the electrodynamics of continuous media independently of a
specific type of the medium. Realistic media are complex systems with a
number of subsystems (free electrons, electrons of different atoms in the
same atomic state, nuclei, etc.). In the macroscopic electrodynamics, after
Lorentz averaging, a medium is considered as continuous substance consisting
of different continuous subsystems. Deformations in media caused by external
forces appear in this approach as a consequence of modifications in particle
trajectories caused by the interaction of those with the external fields.
This means that we can use the methods of the description of the theory of
elasticity \cite{lan2} in the electrodynamics of continuous media. So, the
problem of determination of the four-current of a certain subsystem in an
external field reduces to the problem of determination of the four-current
in the deformed subsystem of the medium. The coordinates of deformed and
undeformed subsystems are connected by Euler transformation in terms of the
displacement field $\mathbf{u(r,}t)$ \cite{lan2}

\begin{equation}
W^{i}\left( X\right) =\left( ct,\mathbf{r-u}\left( \mathbf{r},t\right)
\right)  \tag{15}
\end{equation}
and, hence, the expression (13) which is (10) written for Euler
transformation (15) gives the four-current $j^{i}$ of the subsystem in the
external field as a linear combination of the same quantity $j_{0}^{i}$ for
the undisturbed subsystem. By using (A.5) of Appendix we obtain the
three-dimensional form of the current (13): 
\begin{equation}
\rho \left( \mathbf{r},t\right) =\left\Vert \Lambda \right\Vert \rho
_{0}\left( \mathbf{r-u},t\right) ,  \tag{16.a}
\end{equation}

\begin{equation}
j_{\alpha }\left( \mathbf{r},t\right) =\left\Vert \Lambda \right\Vert
S_{\alpha \beta }^{-1}[\left( \partial _{t}u_{\beta }\right) \rho _{0}\left( 
\mathbf{r-u},t\right) +j_{0\beta }\left( \mathbf{r-u},t\right) ],  \tag{16.b}
\end{equation}
where $\left\Vert \Lambda \right\Vert ,$ $S_{\alpha \beta }^{-1}$ are given
by expression (A.4), (A.6) of Appendix. The expression (16) is exact for any
displacement field. Now, summing the expressions (16) over all subsystems
(each with its own displacement fields) we arrive at the exact formulas for
the charge and current densities of the medium in any external field.

Consider a medium consisting of a subsystem of identical particles and a
background which ensures neutrality of the medium.. Let the particle
distribution in equilibrium be homogeneous and isotropic, then for averaged
charge and current densities we have $\bar{\rho}_{0}=const,$ $\mathbf{\bar{j}%
}_{0}\equiv 0.$ Defining the polarization of the medium as $\mathbf{P}\left( 
\mathbf{r,}t\right) =\rho _{0}\mathbf{u}\left( \mathbf{r,}t\right) $,$%
\mathbf{\ }$which is induced dipole moment density of the particles, and
taking into account the neutrality condition, we obtain from (16), (A.4),
(A.6) the following expressions for the charge and current densities of the
medium in the external field 
\begin{equation}
\bar{\rho}\left( \mathbf{r,}t\right) =-\mathbf{\nabla \cdot \Pi ,} 
\tag{17.a}
\end{equation}

\begin{equation}
\mathbf{\bar{j}}\left( \mathbf{r,}t\right) =\partial _{t}\mathbf{\Pi +}c%
\left[ \mathbf{\nabla \times M}\right] .  \tag{17.b}
\end{equation}

Here

\begin{equation}
\Pi _{\alpha }\equiv P_{\alpha }+\frac{1}{2\rho _{0}}\left[ P_{\nu }\partial
_{\nu }P_{\alpha }-P_{\alpha }\partial _{\nu }P_{\nu }\right] +\frac{1}{%
6\rho _{0}^{2}}e_{\alpha \mu \nu }e_{\beta \lambda \sigma }(\partial _{\mu
}P_{\lambda })(\partial _{\nu }P_{\sigma })P_{\beta },  \tag{18.a}
\end{equation}

\begin{equation}
M_{\alpha }\equiv \frac{1}{2c\rho _{0}}e_{\alpha \lambda \nu }P_{\lambda
}\left( \partial _{t}P_{\nu }\right) +\frac{1}{3c\rho _{0}^{2}}e_{\nu \sigma
\lambda }(\partial _{\alpha }P_{\lambda })\left( \partial _{t}P_{\nu
}\right) P_{\sigma }  \tag{18.b}
\end{equation}
are the electric and magnetic polarization vectors of the medium.

In what follows we will restrict ourselves to consideration of linear regime
of the interaction of external electromagnetic field with isotropic medium,
i.e., we will consider weak electromagnetic fields in an isotropic medium
where $\mathbf{P=}$ $\frac{\varepsilon -1}{4\pi }\mathbf{E.}$This is a
restriction on the value of external electromagnetic field, but it could
have large spatial derivatives which we cannot neglect in this
approximation. Hence, even in linear regime of the interaction expressions
(18) remain nonlinear.

\section{ The field of a point charge in a medium}

\textbf{\ }Consider the electrostatic field of the point particle with
charge $q$ located at the origin of the coordinates in a medium. Because of
spherical symmetry and stationarity of the problem, for the only nonzero
component of (18) we have 
\begin{equation*}
\Pi _{r}=P_{r}-\frac{1}{\rho _{0}}\frac{P_{r}^{2}}{r}+\frac{1}{3\rho
_{0}{}^{2}}\frac{P_{r}^{3}}{r^{2}}.
\end{equation*}
Then the solution to the Maxwell equation for the static field 
\begin{equation*}
\mathbf{\nabla \cdot }\left( \mathbf{E+}4\pi \mathbf{\Pi }\right) =4\pi
q\delta \left( \mathbf{r}\right)
\end{equation*}
in linear nondispersive medium is given by the cubic algebraic equation 
\begin{equation}
E-\frac{\left( \varepsilon -1\right) ^{2}}{\varepsilon (4\pi \rho _{0})}%
\frac{E^{2}}{r}+\frac{\left( \varepsilon -1\right) ^{3}}{\varepsilon (4\pi
\rho _{0})^{2}}\frac{E^{3}}{3r^{2}}=\frac{q}{\varepsilon r^{2}}.  \tag{19}
\end{equation}

Here $\varepsilon $ is the static dielectric constant of the medium. For
large distances from the particle, where $\left\vert u_{r}\right\vert \ll r$
, we may ignore nonlinear terms in (19) and get the well known result:

\begin{equation*}
E=\frac{q}{\varepsilon r^{2}}.
\end{equation*}
But, for small distances, when $\left\vert u_{r}\right\vert \sim r$,
nonlinear corrections in (19) are important. So, for an elementary charge $%
\left\vert q\right\vert =\left\vert e\right\vert $ \ embedded in the
electron gas with the concentration $\sim 10^{23}cm^{-3}$ in solids with $%
\varepsilon =10$ , the corrections are $\sim 1-10\%$ at the distances of a
few atomic radii. These corrections can be important in the problems of an
exciton, solvated electrons and ions, the electronic centers in dye
crystals, etc.

As we see, screening depends on relative sign of $\rho _{0}$ and $q$. For
the same signs of $\rho _{0}$ and $q$ the effective charge is larger than $%
q/\varepsilon $, and for different signs it is smaller. This fact has a
simple geometrical explanation. According to the Gauss theorem, the field of
the point particle at the point $r$ is proportional to the charge within the
sphere of the radius $r.$ In case of the same signs of $\rho _{0}$ and $q$
medium particles between the spheres $r-\left\vert u_{r}\right\vert $ and $r$
leave the sphere $r$. For the case of the different signs of $\rho _{0}$ and 
$q$ medium particles come into the sphere $r$ from between those of radii $r$
and $r+\left\vert u_{r}\right\vert $. But in these two cases the magnitude
of out- and incoming charges are not equal: in the second case the incoming
is greater, and we have the asymmetry mentioned above.

\section{ Nonlinear effects in a linear medium}

\textbf{\ }Let us consider now a problem with cylindrical symmetry, i.e.,
propagation of a radially polarized radiation $\mathbf{E}_{0}\left( r\mathbf{%
,}z,t\right) =\mathbf{\hat{\imath}}_{r}E\left( r\right) \exp i\left(
k_{0}z-\omega _{0}t\right) $ (for instance $TEM_{01}$radially polarized mode 
\cite{pres}) in a linear, isotropic medium. For the nonzero component of $%
\left( 18\right) ,$i.e, radially polarized polarization $\mathbf{P=\hat{%
\imath}}_{r}P_{r}(r,z,t)$ $\left( r^{2}=x^{2}+y^{2}\right) $ we have 
\begin{equation}
\Pi _{r}=P_{r}-\frac{1}{2\rho _{0}}\frac{P_{r}^{2}}{r}.  \tag{20}
\end{equation}
Then (17) and (20) give rise to:

\textbf{a.}Static charge with density

\begin{equation}
\rho \left( r\right) =2\chi ^{\left( 2\right) }\partial _{r}\left\vert
E\left( r\right) \right\vert ^{2},  \tag{21.a}
\end{equation}

\textbf{b.}Charge and current densities at the second harmonic frequency

\begin{equation}
\rho \left( r,z\mathbf{,}t\right) =\chi ^{\left( 2\right) }\partial _{r}%
\mathbf{E}_{0}^{2}\left( r,z\mathbf{,}t\right) ,\text{ \ \ }j_{r}\left( r%
\mathbf{,}z,t\right) =-\chi ^{\left( 2\right) }\partial _{t}\mathbf{E}%
_{0}^{2}\left( r,z\mathbf{,}t\right) ,  \tag{21.b}
\end{equation}

the coefficient of nonlinearity in (20) is 
\begin{equation}
\chi ^{\left( 2\right) }=\left[ \frac{\varepsilon \left( \omega _{0}\right)
-1}{4\pi }\right] ^{2}\cdot \frac{1}{2r\rho _{0}}.  \tag{22}
\end{equation}
\ \ 

So, in a linear medium it is possible to obtain optical detection and second
harmonic generation. Let us estimate coefficient of the quadratic
nonlinearity $\chi ^{\left( 2\right) }$ for the medium with $\varepsilon =3$
and electron concentration $10^{20}\div 10^{21}cm^{-3}.$ For the width of
radiation beam $r\sim 0.1cm$ we have 
\begin{equation*}
\chi ^{\left( 2\right) }\sim 10^{-11}\div 10^{-12}CGSE.
\end{equation*}

In case of focusing this coefficient may reach $\sim 10^{-8}\div 10^{-9}CGSE$
because of strong inhomogeneousity in the focus of the beam; this is large
enough for observation.

\section{\protect\bigskip Concluding remarks}

We have shown how to generate with the help of Einstein covariance principle
infinite many new solutions of a given covariant equation having a known
one. By using Euler transformation we have separated a solution to the
generated set of continuity equation which is a solution to the same
equation in an arbitrary external field. In this way we got an exact
expression for the charge and current densities for a medium in the external
field. This new approach gives us an opportunity to determine the
polarization of a homogeneous medium in a very general manner, i. e.,
independently of the model of the medium. In addition, it is possible, in
this case, to unambiguously determine the electric polarization and
magnetization vectors in terms of the medium polarization and its
derivatives. Further on we considered applications of this expression.
Solving the problem of charged point particle in a homogeneous dielectric
medium we arrived at a cubic algebraic equation for the electric field,
whose solution shows deviation of the electric field from the classical
result at small distances from the point charge. We predicted second
harmonic generation and optical detection phenomena in a linear medium.

It would be interesting to apply this approach to other equations in
mathematical physics. These type of investigations are in progress and will
be published elsewhere.

We note finally that the boundary-value problem of the electrodynamics of an
expanding-contracting sphere has been for the first time solved in \cite{van}
with use of the covariance principle, but the authors did not realized, at
that time, the universality and importance of that approach.

\textbf{Acknowledgements \ }Authors are very grateful to Prof. D.A.
Kirzhnits for stimulating discussions and Profs. R. Balian, V.O. Chaltykyan
and A.O. Melikian for critically reading of the manuscript and for valuable
remarks.

This work was supported by the SCOPES Swiss grant 7UKPJ062150.

\section{\protect\bigskip Appendix: Euler transformation}

\textbf{\ }We use in the paper the Euler transformation which is very well
known in the theory of elasticity:

\begin{equation}
\mathbf{r}^{\prime }\mathbf{=r-u}\left( \mathbf{r,}t\right)   \tag{A.1}
\end{equation}
where $\mathbf{u}\left( \mathbf{r,}t\right) $ is the displacement field of
the medium. \textbf{\ }

For the transformation matrix $\Lambda _{j}^{i}\left( X\right) \equiv \frac{%
\partial X^{\prime i}}{\partial X^{j}}$ and its determinant we get 
\begin{equation}
\Lambda _{0}^{0}=1,\text{\ }\Lambda _{\alpha }^{0}=0,\text{ }\Lambda
_{0}^{\alpha }=-\frac{1}{c}\partial _{t}u_{\alpha },\text{ }\Lambda _{\beta
}^{\alpha }=\delta _{\alpha \beta }-\partial _{\beta }u_{\alpha }\equiv
S_{\alpha \beta },  \tag{A.3}
\end{equation}

\begin{equation}
\left\Vert \Lambda \right\Vert =1-\partial _{\lambda }\left\{ u_{\lambda }+%
\frac{1}{2}\left[ u_{\nu }\partial _{\nu }u_{\lambda }-u_{\lambda }\partial
_{\nu }u_{\nu }\right] +\frac{1}{6}e_{\lambda \mu \nu }e_{\beta \rho \sigma
}(\partial _{\mu }u_{\rho })(\partial _{\nu }u_{\sigma })u_{\beta }\right\} .
\tag{A.4}
\end{equation}

Here $u_{\alpha }$ $\left( \alpha =1,2,3\right) $ are components of $\mathbf{%
u}$. $e_{\alpha \beta \gamma }$, $\delta _{\alpha \beta }$ are three
dimensional Levi-Civita and Kronecker symbols, respectively.

For the reciprocal matrix $\tilde{\Lambda}$ of $\Lambda $ $\left( \Lambda
_{m}^{i}\tilde{\Lambda}_{j}^{m}=\tilde{\Lambda}_{m}^{i}\Lambda
_{j}^{m}=\delta _{j}^{i}\right) $ we find from (A.3):

\begin{equation}
\tilde{\Lambda}_{0}^{0}=1,\text{ }\tilde{\Lambda}_{\alpha }^{0}=0,\text{ }%
\tilde{\Lambda}_{0}^{\alpha }=\frac{1}{c}S_{\alpha \beta }^{-1}\partial
_{t}u_{\beta },\text{ }\tilde{\Lambda}_{\beta }^{\alpha }=S_{\alpha \beta
}^{-1},  \tag{A.5}
\end{equation}
where $\hat{S}^{-1}$ is the reciprocal matrix of $\hat{S}$:

\begin{equation}
\left\Vert \Lambda \right\Vert S_{\alpha \beta }^{-1}=\delta _{\alpha \beta
}-\partial _{\nu }\left[ u_{\nu }\delta _{\alpha \beta }-u_{\alpha }\delta
_{\nu \beta }-\frac{1}{2}u_{\lambda }e_{\lambda \beta \sigma }e_{\nu \alpha
\mu }\partial _{\mu }u_{\sigma }\right] .  \tag{A.6}
\end{equation}

\end{document}